\documentclass[sigconf,screen]{acmart}

\AtBeginDocument{%
	\providecommand\BibTeX{{%
			\normalfont B\kern-0.5em{\scshape i\kern-0.25em b}\kern-0.8em\TeX}}}

 \usepackage{microtype} 
\usepackage{algorithmic}
\usepackage{graphicx}
\usepackage{textcomp}
\usepackage{xcolor}
\usepackage{listings}
\usepackage{enumitem}
\usepackage{booktabs}
\def\BibTeX{{\rm B\kern-.05em{\sc i\kern-.025em b}\kern-.08em
		T\kern-.1667em\lower.7ex\hbox{E}\kern-.125emX}}

\definecolor{codegreen}{rgb}{0,0.6,0}
\definecolor{codegray}{rgb}{0.5,0.5,0.5}
\definecolor{codepurple}{rgb}{0.58,0,0.82}
\definecolor{backcolour}{rgb}{0.95,0.95,0.92}

\lstdefinestyle{mystyle}{
  backgroundcolor=\color{backcolour}, commentstyle=\color{codegreen},
  keywordstyle=\color{magenta},
  numberstyle=\tiny\color{codegray},
  stringstyle=\color{codepurple},
  basicstyle=\ttfamily\footnotesize,
  breakatwhitespace=false,         
  breaklines=true,                 
  captionpos=b,                    
  keepspaces=true,                 
  numbers=left,                    
  numbersep=5pt,                  
  showspaces=false,                
  showstringspaces=false,
  showtabs=false,                  
  tabsize=2
}
\usepackage{subcaption}
\usepackage{cleveref}
\newcommand{\RNum}[1]{\lowercase\expandafter{\romannumeral #1\relax}}
\usepackage{booktabs}

\usepackage{color,soul}
\soulregister\cite7
\soulregister\citep7
\soulregister\ref7
\soulregister\RNum7
\soulregister\item7

\sethlcolor{yellow}

\usepackage{multirow}

\setcopyright{acmlicensed}
\copyrightyear{2025}
\acmYear{2025}
\acmDOI{XXXXXXX.XXXXXXX}
\acmConference[Conference acronym 'XX]{Make sure to enter the correct
  conference title from your rights confirmation email}{June 03--05,
  2018}{Woodstock, NY}

\begin{document}
	
	\title{Evaluating Large Language Models for Code Translation: Effects of Prompt Language and Prompt Design}

\author{Aamer Aljagthami, Mohammed Banabila, Musab Alshehri, Mohammed Kabini, and Mohammad D. Alahmadi}
\affiliation{%
    \institution{Department of Software Engineering, College of Computer Science and Engineering, University of Jeddah}
    \city{Jeddah}
    \postcode{21493}
    \country{Saudi Arabia}
}
\email{{aaljagthami0002, mbanabeelah, malshehri0625, mkabini, mdalahmadi}@uj.edu.sa}

	
	\begin{abstract}









Large language models (LLMs) have shown promise for automated source-code translation, a capability critical to software migration, maintenance, and interoperability. Yet comparative evidence on how model choice, prompt design, and prompt language shape translation quality across multiple programming languages remains limited. This study conducts a systematic empirical assessment of state-of-the-art LLMs for code translation among C++, Java, Python, and C\#, alongside a traditional baseline (TransCoder). Using BLEU and CodeBLEU, we quantify syntactic fidelity and structural correctness under two prompt styles (concise instruction and detailed specification) and two prompt languages (English and Arabic), with direction-aware evaluation across language pairs. Experiments show that detailed prompts deliver consistent gains across models and translation directions, and English prompts outperform Arabic by 13--15\%. The top-performing model attains the highest CodeBLEU on challenging pairs such as Java$\rightarrow$C\# and Python$\rightarrow$C++. Our evaluation shows that each LLM outperforms TransCoder across the benchmark. These results demonstrate the value of careful prompt engineering and prompt language choice, and provide practical guidance for software modernization and cross-language interoperability.

	\end{abstract}

	\keywords{
    Code Translation, LLMs, BLEU, CodeBLEU.
	}
	
	\maketitle

	\section{Introduction}
	\label{sec:introduction}

Large language models (LLMs) have shown promise for automated source code translation, a capability critical to software migration, maintenance, and interoperability \cite{eniser2024towards}. Translating across languages can reduce technical debt, enable reuse, and support polyglot development workflows \cite{niephaus2018live}. At the same time, code translation remains challenging due to differences in type systems, idioms, libraries, and runtime models \cite{zhang2025scalable}.

Despite rapid progress, comparative evidence on LLM-based code translation is still limited. Prior studies often vary in datasets, metrics, and evaluation setups, which hinders fair comparison \cite{jiao2023evaluation}. The roles of prompt design and prompt language are not well quantified across models and directions. Direction-aware analyses are also scarce, even though translation difficulty can differ markedly from source to target \cite{tao2024unraveling}. Baseline comparisons to traditional systems such as TransCoder are likewise inconsistent \cite{jiao2023evaluation}.

This paper conducts a systematic empirical assessment of LLMs for code translation among C++, Java, Python, and C\#, with a side-by-side comparison to a traditional baseline under a unified setup. We measure performance with BLEU and CodeBLEU to capture both surface correspondence and structural correctness \cite{ren2020codebleu}. We study two prompt styles (concise instruction and detailed specification) and two prompt languages (English and Arabic), and we use direction-aware evaluation across language pairs. Our experiments show three consistent patterns: detailed prompts yield measurable gains across models and directions, English prompts outperform Arabic by roughly 13--15\% on average, and each LLM surpasses the baseline under identical evaluation settings. These results demonstrate the value of careful prompt engineering and prompt language choice, and they provide actionable guidance for software modernization and cross-language interoperability.

Our contributions are summarized as follows:
\begin{itemize}[itemindent=1em]
\item We conduct a systematic empirical assessment of LLMs for code translation among C++, Java, Python, and C\#, with a side-by-side comparison to a traditional baseline under identical evaluation settings.
\item We study the effects of prompt design by contrasting concise instructions with detailed task specifications, and we quantify the impact of prompt language by evaluating English and Arabic variants.
\item We provide direction-aware analyses across language pairs to reveal asymmetries that affect translation difficulty and model behavior.
\item We report comprehensive results with BLEU and CodeBLEU, and derive practical recommendations for deploying LLMs in automated code migration workflows, and provide a replication package\footnote{https://doi.org/10.5281/zenodo.17065727}
\end{itemize}

Our experiments indicate three consistent patterns. First, detailed prompts deliver measurable gains across models and translation directions. Second, English prompts outperform Arabic by approximately 13-15\% on average. Third, all evaluated LLMs exceed the TransCoder baseline under the same conditions, with the strongest CodeBLEU observed on challenging directions such as Java$\rightarrow$C\# and Python$\rightarrow$C++. These findings demonstrate the value of careful prompt engineering and prompt language choice, and they offer actionable guidance for practitioners seeking reliable, high-quality code translation at scale.










	\section{Related Work}
	\label{sec:related_work}








\subsection{Classical and Neural Code Translation}
Early systems approached code translation with rule-based transpilers and compiler-driven rewriting, which ensured syntactic fidelity but struggled with idioms, library mappings, and semantic gaps across ecosystems \cite{aggarwal2015using}. Neural machine translation reframed the task as sequence-to-sequence learning over tokens or subwords, sometimes augmented with abstract syntax trees or grammar constraints to improve structural validity \cite{ahmed2005syntax}. Large-scale pretraining for code, as in PLBART and CodeT5, improved downstream translation and repair by leveraging mined repositories and task-adaptive finetuning \cite{wang2021codet5}. TransCoder demonstrated that denoising objectives and mined parallel data provide strong baselines, although performance varied by direction and often required post-processing \cite{roziere2020unsupervised}.

\subsection{LLM-based Translation, Prompting, and Comparability}
Recent large language models trained on mixed natural language and code have advanced program synthesis, editing, and translation without task-specific finetuning \cite{sagodi2024methodology} \cite{yin2024rectifier}. Prompt design has emerged as a central factor, including explicit constraints, structured I/O formats, and function-signature hints \cite{fu2025first}. Multilingual prompting raises additional questions about the effect of prompt language on model behavior, yet controlled comparisons remain limited in code translation \cite{ghosh2025multilingual}. Our study addresses these gaps by using a unified evaluation protocol across four languages (C++, Java, Python, C\#), by contrasting concise versus detailed prompts and English versus Arabic prompts, and by reporting direction-aware analyses that expose asymmetries between source and target.

Compared with prior work, we provide a head-to-head comparison of contemporary LLMs against a traditional baseline under identical settings, isolate the effects of prompt style and prompt language, and present results stratified by translation direction. This design improves comparability and yields practical guidance for software modernization and cross-language interoperability.

\section{Empirical Study}
\label{sec:study}

This section states the research questions, describes the datasets and tasks, and details the models, prompting protocol, metrics, and procedure used in our evaluation. The goal is a controlled and comparable assessment of LLM-based code translation for software modernization and cross-language interoperability.

\subsection{Research Questions}
\label{subsec:rqs}
We organize the study around three questions:
\begin{itemize}[itemindent=1em]
  \item[\textbf{\textit{RQ$_1$}}] \textit{Model choice:} Which state-of-the-art LLM achieves the highest translation quality across directed language pairs, and how does it compare with a traditional baseline under identical settings?
  \item[\textbf{\textit{RQ$_2$}}] \textit{Prompt factors:} How do prompt style (concise instruction versus detailed specification) and prompt language (English versus Arabic) affect translation quality across models and directions?
  \item[\textbf{\textit{RQ$_3$}}] \textit{Directionality:} How does performance vary by translation direction among C++, Java, Python, and C\#, and what asymmetries emerge between source and target languages?
\end{itemize}

\subsection{Datasets}
\label{subsec:datasets}

The dataset we used for our experiment comes from the MuST (Multilingual Code Snippets Training for Program Translation) dataset \cite{zhu2022multilingual}, a publicly available benchmark designed for assessing code translation models across multiple programming languages. This dataset comprises parallel implementations (same functions in the same files for the various languages) of functions in C++, Java, Python, and C\#, ensuring direct comparability between equivalent code snippets in different languages.

For this study, 24 functions (6 per language) were randomly sampled from the dataset to maintain diversity and mitigate selection bias. The way to randomly sample the dataset was to use a random number generator between 1 and 2702 to choose between the 2702 entries. Entries where the code did not include all of the four languages specified in the study were ignored. Each function was paired with its corresponding translations in the other languages, serving as ground-truth references for evaluation. For illustration, Listings 1 and~\ref{2nd} show a Python–Java pair that implements the same recursive check for identical linked lists; the Java version separates the recursion into a helper method to follow standard object-oriented conventions.

\begin{lstlisting}[language=Python,  caption=Python example]
def areIdentical(a, b):
     
    if (a == None and b == None):
        return True
 
    if (a != None and b != None):
        return ((a.data == b.data) and
                 areIdentical(a.next, b.next))
 
    return False
\end{lstlisting}

\begin{lstlisting}[language=Java, label=2nd, caption=Java example]
boolean areIdenticalRecur(Node a, Node b)
{

    if (a == null && b == null)
        return true;
 
    if (a != null && b != null)
        return (a.data == b.data) &&
               areIdenticalRecur(a.next, b.next);
 
    return false;
}
 
boolean areIdentical(LinkedList listb)
{
    return areIdenticalRecur(this.head, listb.head);
}

\end{lstlisting}

\subsection{Models and Baseline}
\label{subsec:models}
We compare four state-of-the-art LLMs for code translation—GPT-4o (OpenAI), Gemini 2.0 (Google), DeepSeek-V3 (DeepSeek AI), and Claude 3.7 (Anthropic)—in a zero-shot setting (no fine-tuning), and we include TransCoder as a traditional baseline \cite{zhu2022multilingual}.

\subsection{Prompt Design}
\label{subsec:prompt_protocol}
We study two prompt styles and two prompt languages.

\paragraph{Concise prompts.}
For each directed pair we issue a short instruction. Examples:  
C++ $\rightarrow$ Java: “Please translate this code from C++ to Java.”;  
C++ $\rightarrow$ Python: “Please translate this code from C++ to Python.”;  
C++ $\rightarrow$ C\#: “Please translate this code from C++ to C\#.”  
Similarly for Java, Python, and C\# as sources: “Please translate this code from \emph{X} to \emph{Y}.” We instantiate each instruction in English and in Arabic.

We batch prompts using a CSV with four input columns: \emph{Code\_ID}, \emph{Source\_Language}, \emph{Target\_Language}, \emph{Source\_Code}. The model is instructed to translate from the source to the target language under the following requirements: (1) preserve logic, algorithm, and computational flow; (2) maintain variable names where syntactically valid; (3) keep function or method structure and naming conventions adapted to the target language; (4) preserve comments and their relative positions; (5) maintain control-flow structure; (6) use equivalent data structures and types; (7) ensure identical outputs for identical inputs; (8) follow target-language naming conventions; (9) include necessary imports or includes; (10) preserve code organization and hierarchy. The model is asked to return only code unless otherwise requested. The output is written to a CSV with six columns: \emph{Code\_ID}, \emph{Source\_Language}, \emph{Target\_Language}, \emph{Source\_Code}, \emph{Translated\_Code}, \emph{Status}. When rules (2) and (8) conflicted, models were expected to prioritize target language conventions while preserving semantic meaning.

\subsection{Evaluation Metrics}
\label{subsec:metrics}
We report BLEU and CodeBLEU to capture surface correspondence and structural similarity, respectively \cite{ren2020codebleu}. BLEU measures n-gram overlap with references. CodeBLEU augments n-gram matching with syntax- and data-flow–aware components. We present direction-aware scores as well as macro-averages to surface asymmetries between source and target languages.

\section{Results and Discussion}
\label{sec:results}

\subsection{RQ\textsubscript{1}: Model Choice}
\label{subsec:rq1_models}
\begin{figure}[t]
    \centering
    \includegraphics[width=\linewidth]{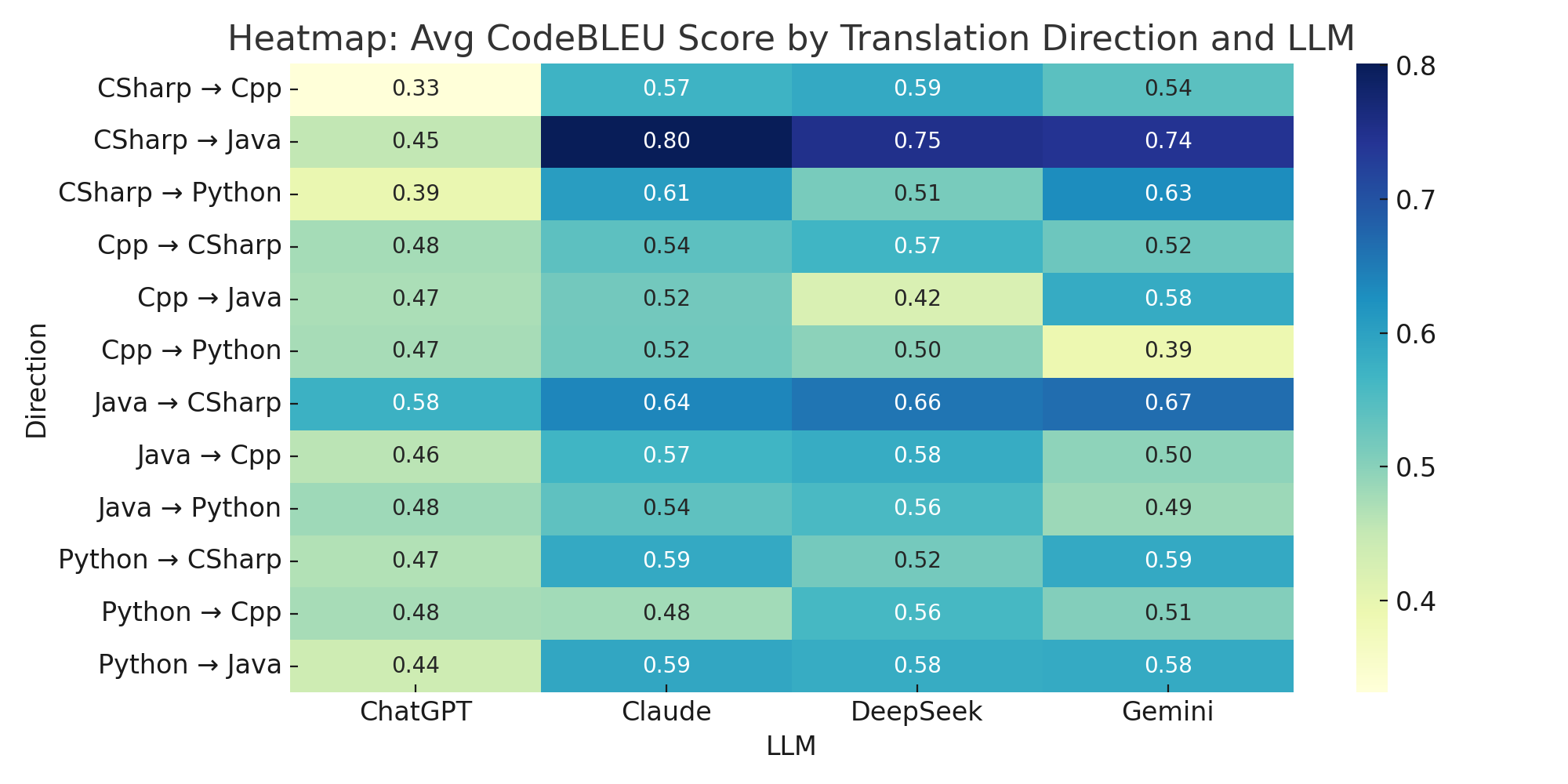}
    \caption{CodeBLEU by model and direction using the simple prompt.}
    \label{fig:codebleu_heatmap_simple}
\end{figure}

\begin{figure}[t]
    \centering
    \includegraphics[width=\linewidth]{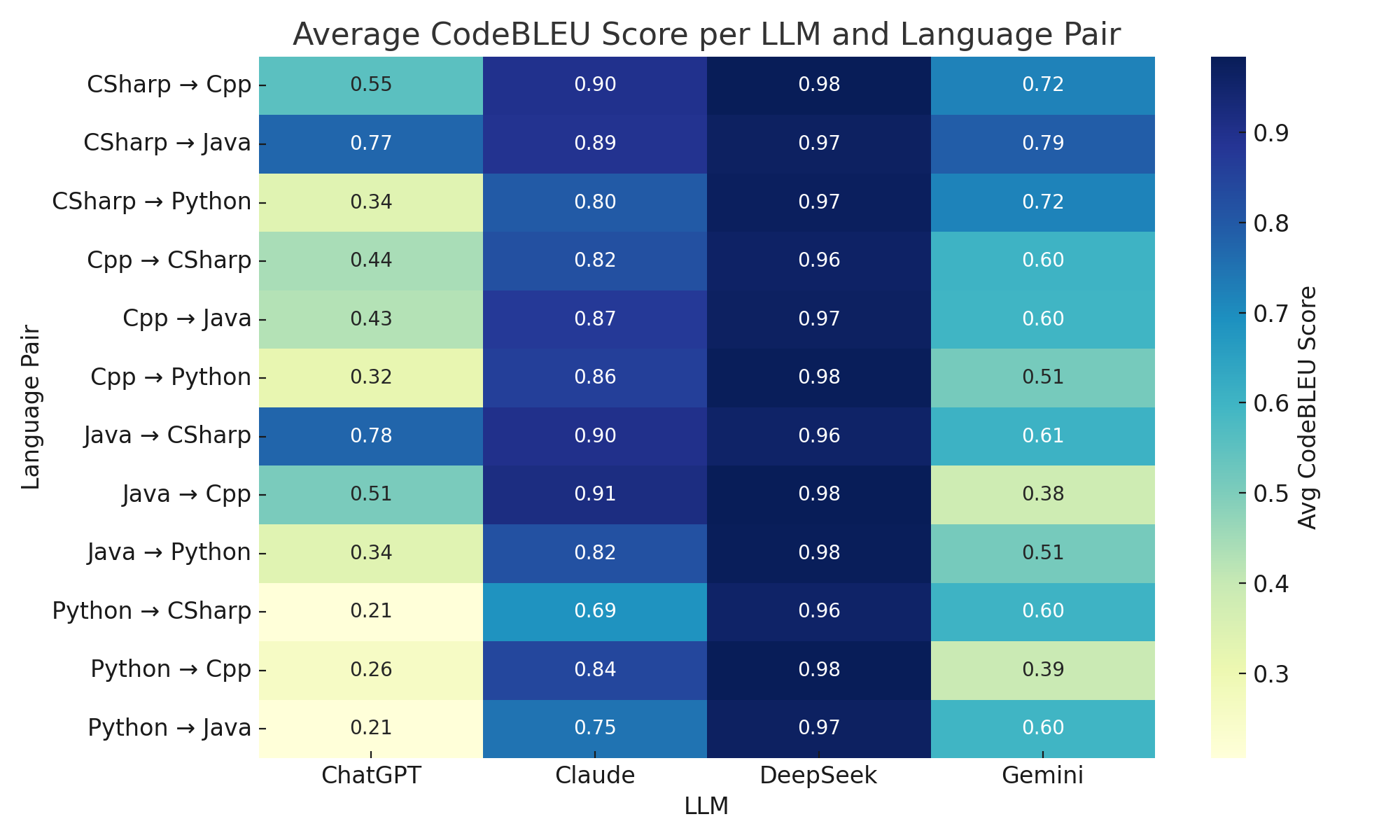}
    \caption{CodeBLEU by model and direction using the detailed prompt.}
    \label{fig:codebleu_heatmap_detailed}
\end{figure}

\begin{figure}[t]
    \centering
    \includegraphics[width=\linewidth]{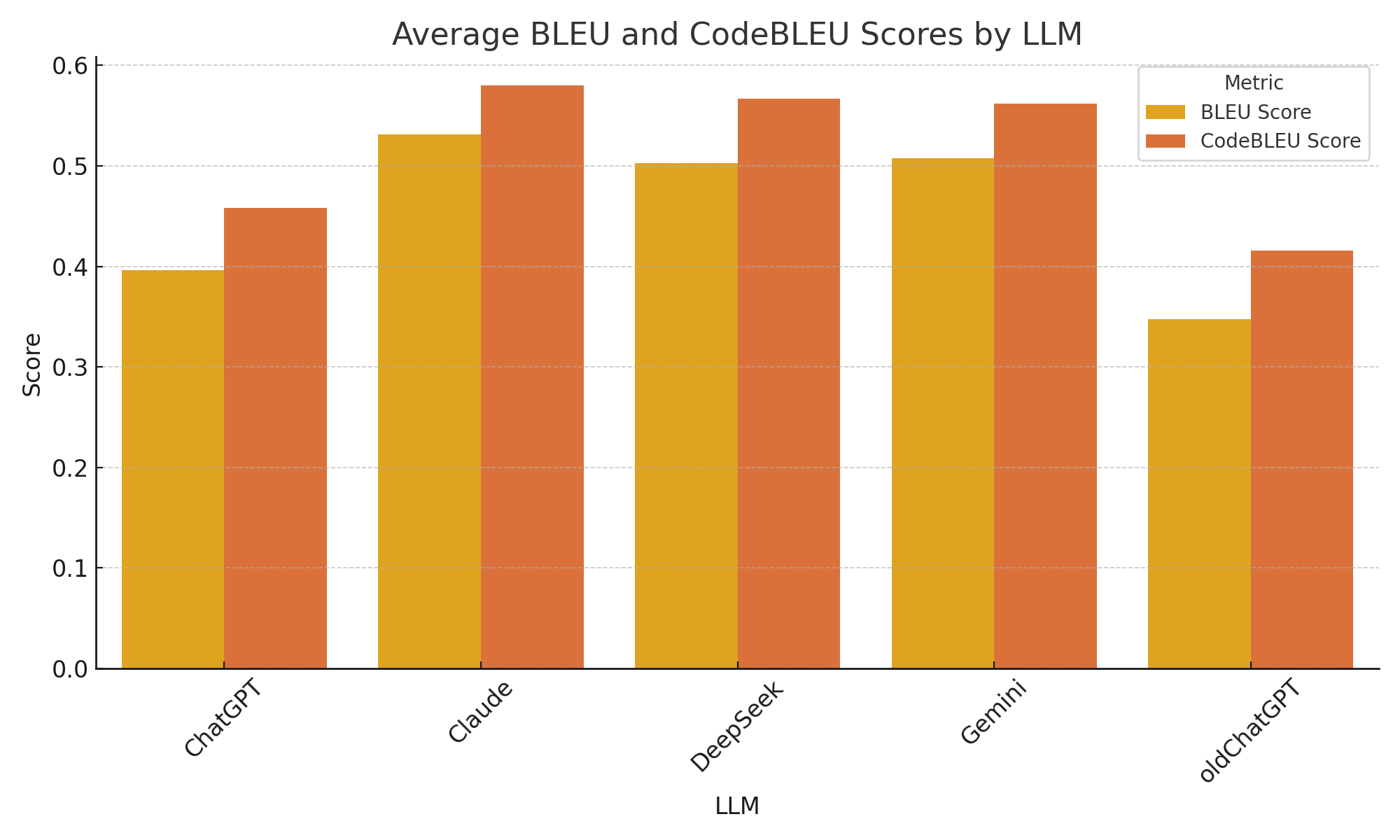}
    \caption{BLEU and CodeBLEU per LLM under the simple prompt.}
    \label{fig:llm_bleu_codebleu_simple}
\end{figure}

DeepSeek and Claude achieve the strongest overall translation quality across directions. DeepSeek leads on challenging pairs such as Java$\rightarrow$C\# and Python$\rightarrow$C++ in the CodeBLEU heatmaps (Figs.~\ref{fig:codebleu_heatmap_simple}, \ref{fig:codebleu_heatmap_detailed}). GPT-4o is competitive on several pairs but trails the top two on stricter targets; Gemini generally ranks lower. \textbf{Table~\ref{tab:my-table}} (BLEU) reinforces this ordering and shows a wide gap between LLMs and traditional baselines across most directions. As shown in Fig.~\ref{fig:llm_bleu_codebleu_simple}, the per-model BLEU and CodeBLEU under the simple prompt follow the same ordering observed in the heatmaps.

\subsection{RQ\textsubscript{2}: Prompt Factors (Style and Language)}
\label{subsec:rq2_prompts}
\begin{figure}[t]
    \centering
    \includegraphics[width=\linewidth]{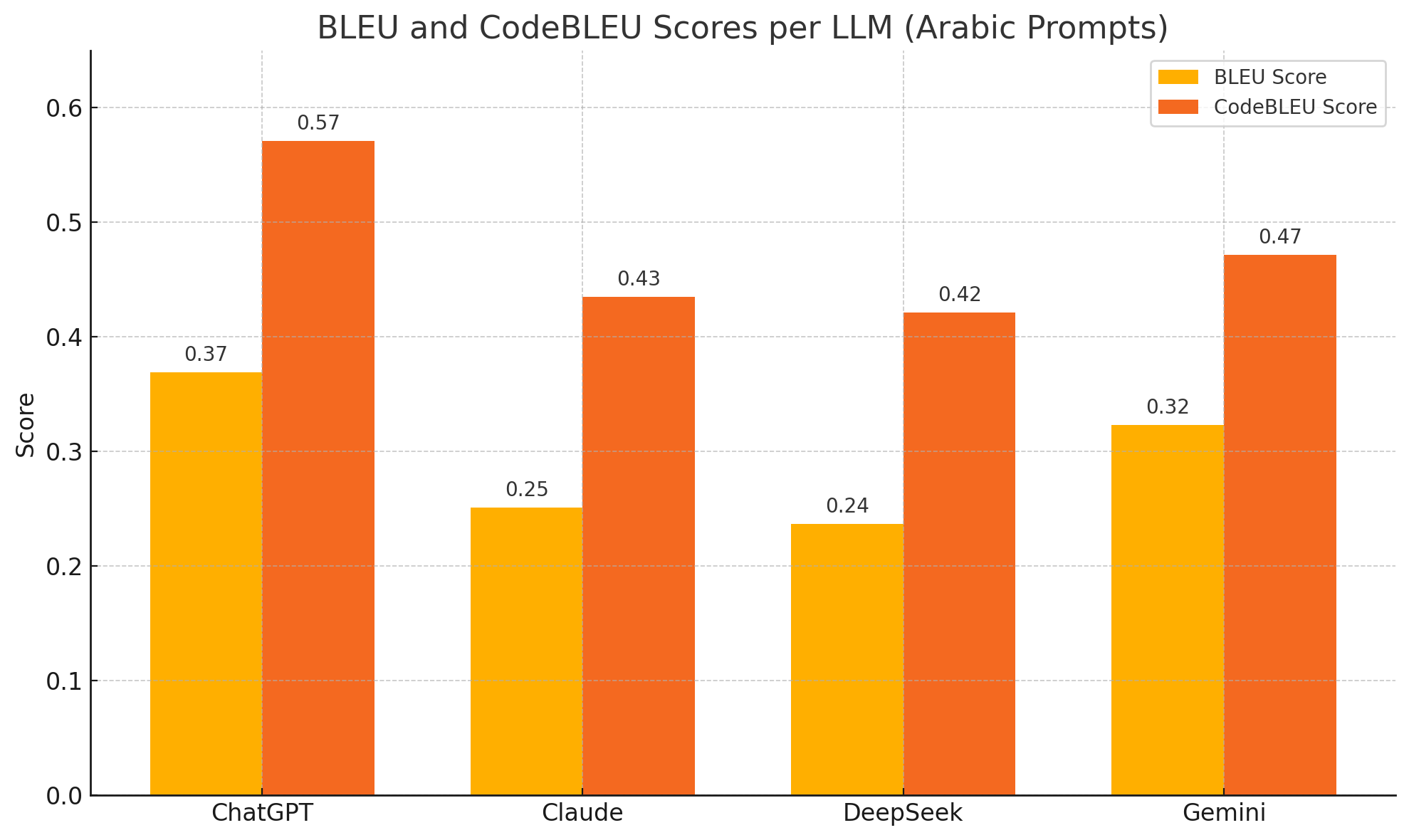}
    \caption{English vs.\ Arabic prompts (BLEU and CodeBLEU).}
    \label{fig:prompt_lang_bleu_codebleu}
\end{figure}

Prompt structure and prompt language both affect outcomes. Comparing the simple vs.\ detailed prompts (Figs.~\ref{fig:codebleu_heatmap_simple}, \ref{fig:codebleu_heatmap_detailed}), the detailed specification yields consistent gains across models and directions, amplifying DeepSeek’s lead over GPT-4o and Gemini. Prompt language also matters: English prompts outperform Arabic by about 13--15\% in CodeBLEU, with similar trends in BLEU (Fig.~\ref{fig:prompt_lang_bleu_codebleu}).

\subsection{RQ\textsubscript{3}: Directionality}
\label{subsec:rq3_direction}
\begin{figure}[t]
    \centering
    \includegraphics[width=\linewidth]{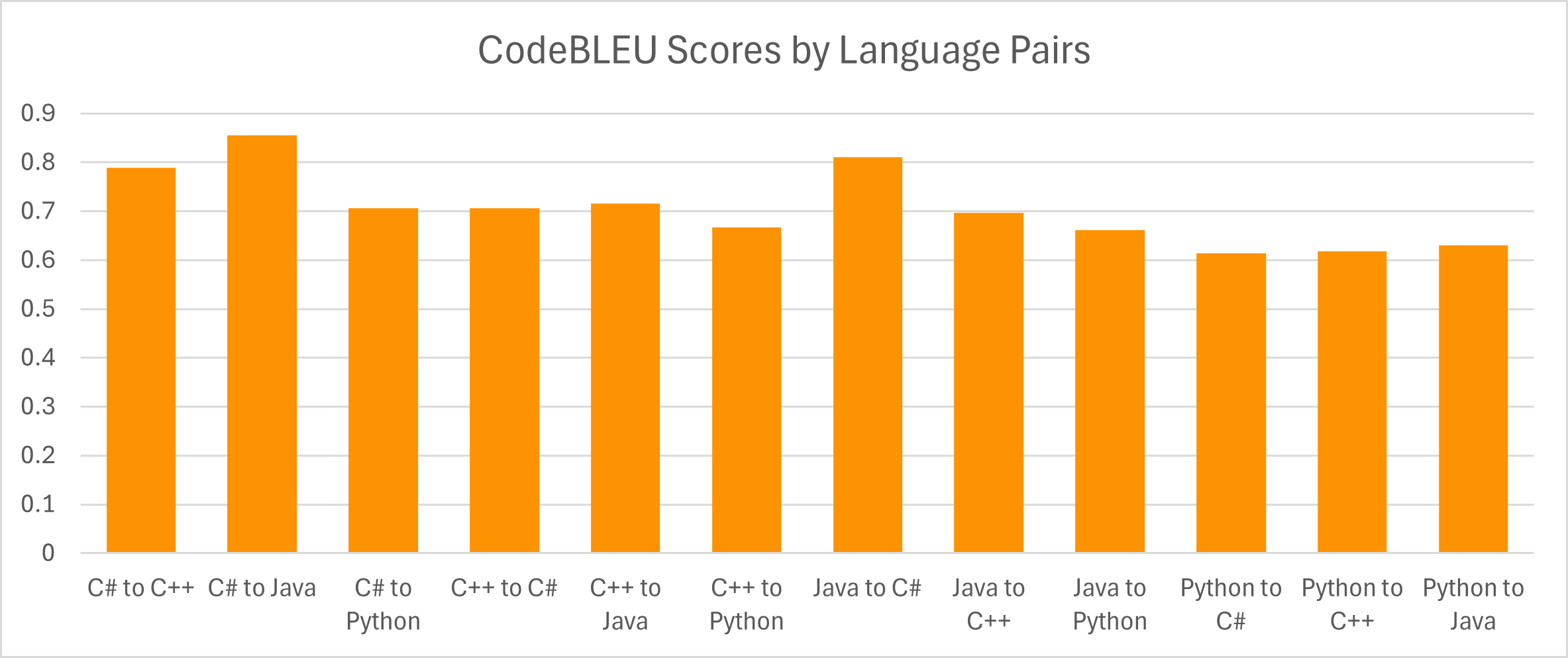}
    \caption{Direction-wise CodeBLEU under the detailed prompt.}
    \label{fig:codebleu_pairs_detailed}
\end{figure}

Performance is direction-dependent. Under the detailed prompt (Fig.~\ref{fig:codebleu_pairs_detailed}), we observe asymmetries such as C\#$\rightarrow$Java $>$ Java$\rightarrow$C\#, and Python$\rightarrow$Java $>$ Java$\rightarrow$Python across most models. Directions into stricter type systems (e.g., $\cdot\rightarrow$C\# or $\cdot\rightarrow$C++) are typically harder than into Python.

\begin{table*}[t]
\centering
\resizebox{\textwidth}{!}{%
  \begin{tabular}{@{}llllllllllllll@{}}
  \toprule
  \textbf{Source} & \textbf{LLM} & \textbf{Java-Py} & \textbf{Py-Java} &
  \textbf{Java-C++} & \textbf{C++-Java} & \textbf{Java-C\#} & \textbf{C\#-Java} &
  \textbf{Py-C++} & \textbf{C++-Py} & \textbf{Py-C\#} & \textbf{C\#-Py} &
  \textbf{C++-C\#} & \textbf{C\#-C++} \\
  \midrule
  Baseline & Naive Copy       & 34.56 & 34.27 & 66.53 & 66.57 & 77.15 & 77.23 & 36.58 & 36.58 & 35.69 & 35.76 & 67.22 & 67.16 \\
  Baseline & Transformer      & 31.22 & 38.15 & 44.38 & 43.93 & 47.34 & 45.6  & 37.42 & 33.9  & 36.91 & 32.64 & 45.32 & 42.65 \\
  Baseline & Transformer+MuST & 40.9  & 43.97 & 58.35 & 54.61 & 73.7  & 71.68 & 42.86 & 39.06 & 43.42 & 42.34 & 57.84 & 57.49 \\
  Baseline & CodeBERT         & 38.7  & 41.35 & 65.48 & 53.47 & 85.46 & 82.45 & 43.96 & 38.37 & 46.4  & 41.1  & 63.01 & 67.17 \\
  Baseline & CodeBERT+MuST    & 55.5  & 57.66 & 81.09 & 78.69 & 90.47 & 86.76 & 58.91 & 55.98 & 59.13 & 55.45 & 79.05 & 81.54 \\
  Baseline & TransCoder       & 24.98 & 21.98 & 30.09 & 30.42 & 44.85 & 29.4  & 23.03 & 23.52 & 40.4  & 18.81 & 41.91 & 25.3  \\
  Baseline & TransCoder+MuST  & 60.73 & 65.53 & 87.09 & 81.64 & 91.74 & 27.7  & 68.7  & 62.92 & 66.52 & 16.88 & 82.4  & 29.44 \\
  LLM      & Claude           & 81.68 & 74.81 & 91.39 & 87.22 & 90.17 & 89.17 & 84.13 & 86.00 & 68.82 & 79.66 & 82.04 & 89.64 \\
  LLM      & Gemini           & 51.09 & 59.62 & 38.48 & 59.62 & 60.67 & 79.12 & 39.25 & 51.09 & 60.21 & 71.67 & 60.21 & 72.00 \\
  LLM      & ChatGPT          & 33.75 & 20.74 & 50.59 & 42.54 & 77.60 & 77.39 & 25.59 & 31.83 & 20.61 & 33.71 & 44.25 & 55.40 \\
  LLM      & DeepSeek         & 97.99 & 96.69 & 98.37 & 96.69 & 95.76 & 96.69 & 98.14 & 97.99 & 95.76 & 97.42 & 96.00 & 98.37 \\
  \bottomrule
  \end{tabular}%
} 
\newline
\newline
\caption{BLEU Scores for LLMs compared with models from the Multilingual Code Snippets Training for Program Translation study.}
\label{tab:my-table}
\end{table*}

\section{Conclusion}
\label{sec:conclusion}

This study evaluated state-of-the-art LLMs for code translation among C++, Java, Python, and C\#, using a MuST-based test set and measuring quality with BLEU and CodeBLEU under two prompt styles and two prompt languages. Three findings stand out. First, model choice matters: DeepSeek and Claude deliver the strongest overall performance, with DeepSeek leading on challenging directions such as Java$\rightarrow$C\# and Python$\rightarrow$C++. GPT-4o is competitive on several pairs, while Gemini generally ranks lower. Second, prompt factors are influential: detailed task specifications yield reliable gains over concise instructions, and English prompts outperform Arabic by about 13--15\%. Third, translation is direction-dependent: we observe stable asymmetries (for example, C\#$\rightarrow$Java exceeds Java$\rightarrow$C\#, and Python$\rightarrow$Java exceeds Java$\rightarrow$Python), with targets that enforce stricter type systems typically harder.

These results provide practical guidance for software modernization and cross-language interoperability. Practitioners should prefer detailed specifications over minimal instructions, use English prompts when feasible, report direction-aware scores rather than only averages, and select models with attention to target-language strictness. Future work will explore scaling the benchmark, adding execution-based checks and human assessment, broadening language coverage, and developing prompt designs or tuning strategies that improve performance with non-English prompts.

\end{document}